\documentclass[runningheads]{template/llncs}
\usepackage{amsmath}
\usepackage{algorithm}
\usepackage{algpseudocode}
\usepackage{quantikz}

\usepackage{algorithm}
\usepackage{algpseudocode}

\usepackage{epsfig, amsmath, amsfonts, tikz, url}
\usetikzlibrary{quantikz2} % in tikz for quantum 

\usepackage{caption}
\usepackage{subcaption}
\usetikzlibrary{matrix, decorations}
\usepackage{amssymb}
\usepackage{multirow}
\usepackage{bbm}
\usepackage{mathrsfs}
\usepackage{cancel}% for strikethrough in math
\usepackage[normalem]{ulem} % for strikesthrough of text
\usepackage{breqn} %break lines in \begin{dmath} for long formula
\usepackage{esvect} % for vector \vv{}
\usepackage{booktabs}
\usepackage{float}
\usepackage{diagbox}
\usepackage{hyperref}

\usepackage{lipsum}

\usepackage[shortlabels]{enumitem}

\usepackage{mathtools}
\def\leftsemijoin{\mbox{$\mathrel{\raise1pt\hbox{\vrule height6pt depth0pt width0.6pt\hskip-1.5pt$>$\hskip -2.5pt$<$}}$}}
\def\rightsemijoin{\mbox{$\mathrel{\raise1pt\hbox{\hskip-1.5pt$>$\hskip -2.5pt$<$\hskip -1.5pt\vrule height6pt depth0pt width0.6pt}}$}}

\makeatletter
\title{Exploring Quantum Bootstrap Sampling for AQP Error Assessment: A Pilot Study}
\author{
Feng Yu 
\and
Raya Jahan 
}
\institute{
  Department of Computer Science and Information Systems\\
  Youngstown State University, Youngstown, OH, USA\\
  \email{fyu@ysu.edu, rjahan01@student.ysu.edu}
 }

\let\Title\@title
\makeatother

\begin{document}
\maketitle
% %%---latex--- 
\begin{abstract}
Error assessment for Approximate Query Processing (AQP) is a challenging problem. Bootstrap sampling can produce error assessment even when the population data distribution is unknown. However, bootstrap sampling needs to produce a large number of resamples with replacement, which is a computationally intensive procedure. In this paper, we introduce a quantum bootstrap sampling (QBS) framework to generate bootstrap samples on a quantum computer and produce an error assessment for AQP query estimations. The quantum circuit design is included in this framework.
\end{abstract}

\section{Introduction}

% - aqp
% - error assessment
% - bootstrap sampling
% - quantum computing
% - contribution
%     - intro qbs
%     - develop q sampling with replacement
%     - design circuit
% - paper structure

In the era of big data, computing complex queries requires a tremendous amount of time, posing emerging challenges in many computing fields. Approximate Query Processing (AQP) \cite{li-tods2019,Ling1999,encyc-aqp} aims to reduce the original big data into small data by employing sampling methods and statistical estimators to provide prompt and accurate approximated query answers.

A well-known problem for AQP research is to estimate the error of a query estimation. The existing methods in this research usually require strict assumptions on the original dataset. Bootstrap sampling \cite{efron94bootstrap,yu2022non} is a nonparametric statistical method that can produce estimations of the error for a given AQP answer even when the statistics of the original dataset are unknown. However, bootstrap sampling requires producing a large number of bootstrap resamples, which are random samples with replacement, based on the sample data, or sample synopses, used by AQP methods. After that, estimations on the bootstrap resamples shall be produced, named bootstrap replications. This procedure can be computationally intensive when a large number of bootstrap samples is needed, and even worse when the sample dataset is also large.

Different from classical computers, which use classical bits, quantum computers are developed to compute based on quantum bits, or qubits. Unlike classical bits, which can only be either 0 or 1 at a time, a qubit can be both $\ket{0}$ and $\ket{1}$ at the same time, named a superposition. The superpositions are employed by a quantum computer to achieve quantum parallelism to tackle computing-intensive problems. For example, the large number factoring problem, which forms the security backbone of modern cryptographic methods, such as RSA, can be promptly solved by Shor's algorithm\cite{wang2012quantum} on a quantum computer.

The contribution of this research is threefold. First, we explore using a quantum computer to implement bootstrap resampling with replacement on a given sample of data. Second, we introduce quantum bootstrap sampling (QBS), a bootstrap sampling framework used for AQP error assessment implemented using quantum computing. Finally, we design the quantum circuit to implement the introduced quantum bootstrap sampling framework. Experimental results show that the proposed quantum circuit can correctly produce desired output.

The paper is structured as follows. Section \ref{sec:background} includes the background. Section \ref{sec:problem} includes the problem statement. Section \ref{sec:methodology} introduces the framework of quantum bootstrap sampling. We present experimental results in Section \ref{sec:experiment}. The related work is included in Section \ref{sec:related-work}. The conclusion and future work are included in Section \ref{sec:conclusion}.

\section{Background} \label{sec:background}

% - bootstrap sampling
% - error assessment formula
% - quantum computing
% - quantum parallel

Bootstrap sampling is a non-parametric statistical method often employed when error assessment is required for an estimator based on a statistical sample. A unique procedure in bootstrap sampling is \textit{resampling}, which will generate many new random samples with replacement, named bootstrap samples, from the given sampled dataset. Usually, the larger the number of bootstrap samples generated, the better the accuracy of the error assessment. However, this process can be computationally intensive if many bootstrap samples are needed.

Given a sample $\vec{y}=(y_1, y_2, ..., y_n)$ from an unknown distribution $F$, a bootstrap sample $\vec{y}^*=(y^*_1, y^*_2, ..., y^*_n)$ is a resampled collection obtained by randomly sampling $n$ times with replacement from $\vec{y}$. For example, if $n=5$, we might obtain various bootstrap samples with different combinations such as $\vec{y}^*_1=(y_5, y_3, y_1, y_2, y_1)$, $\vec{y}^*_2=(y_2, y_5, y_4, y_1, y_2)$, $\vec{y}^*_3=(y_3, y_3, y_2, y_3, y_4)$ etc. After summarizing the frequency of each element sampled, we obtain the distribution of a bootstrap sample, $\hat{F}=(\hat{f}_1, \hat{f}_2, \dots)$, where $\hat{f}_k=\#\{y^*_i=y_k\}/n$. 

\begin{figure}
    \centering
    \begin{subfigure}[t]{0.4\textwidth}
        \centering        \includegraphics[width=1\textwidth]{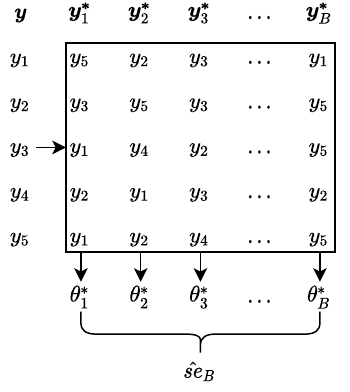}
        \caption{Bootstrap Samples}
        \label{fig:bs:eg1}
    \end{subfigure}
    \begin{subfigure}[t]{0.4\textwidth}
        \centering
        \includegraphics[width=1\textwidth]{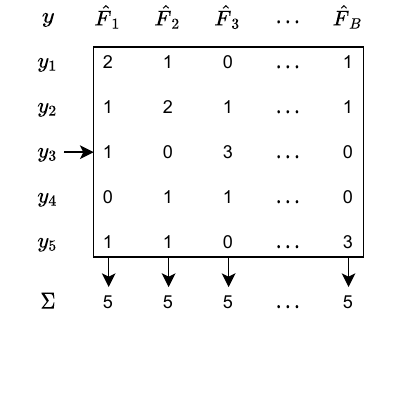}
        \caption{Bootstrap Distributions}
        \label{fig:bs:eg2}
    \end{subfigure}
\vspace{-6pt}
\caption{Example: Bootstrap Sampling}
\label{fig:bs:eg}
\end{figure}

Figure \ref{fig:bs:eg} depicts the examples of bootstrap samples and bootstrap distribution.

A typical application of bootstrap sampling is to estimate the standard error of a sample estimator even when the population distribution is unknown. Suppose that we are interested in a parameter $\theta=t(F)$ calculated based on $\vec{y}$. Usually, we do not have all the information of $F$, but can only calculate an estimation of $\theta$ from the given sample $\vec{y}$ denoted by $\hat{\theta}=s(\vec{y})$. For each bootstrap sample $\vec{y}^*$ we can generate a \emph{bootstrap replication} of $\hat{\theta}$, denoted by

$$
\hat{\theta}^*=s(\vec{y}^*)
$$
For example, when $\hat{\theta}$ is the sample mean $\overline{\vec{y}}$, a bootstrap replication $\hat{\theta}^*$ is the sample mean on a bootstrap sample $\overline{\vec{y}^*}$.

After generating several $B$ independent bootstrap samples, we can obtain the standard error of all $\hat{\theta}^*$, i.e. $\widehat{se}_B(\hat{\theta}^*)$, called the \textit{bootstrap estimation of the standard error}. When $B\rightarrow \infty$, $\widehat{se}_B(\hat{\theta}^*) \rightarrow se_{\hat{F}}(\hat{\theta}^*)$.

$se_{\hat{F}}(\hat{\theta}^*)$ is called the \emph{ideal bootstrap estimation} of the ground truth standard error of $\hat{\theta}$, $se_{F}(\hat{\theta})$. We say that $se_{\hat{F}}(\hat{\theta}^*)$ is a \emph{plug-in} estimate of $se_{F}(\hat{\theta})$ that uses the empirical distribution $\hat{F}$ in replacement of the unknown distribution $F$.

$\widehat{se}_B(\hat{\theta}^*)$ can be calculated as,

\begin{equation}\label{eq:bs:se}
\widehat{se}_B(\hat{\theta}^*) = \left[\frac{1}{B-1}\sum_{i=1}^{B} \left( \hat{\theta}^*(i) - \bar{\theta}^*\right)^2  \right]^{\frac{1}{2}}
\end{equation}
where $\bar{\theta}^*=\sum^B_{i=1}\hat{\theta}^*(i)/B$.

% Both $se_{\hat{F}}(\hat{\theta}^*)$ and its approximation $\widehat{se}_B(\hat{\theta}^*)$ are called \emph{non-parametric bootstrap} estimates since they are generated from the distributions, $\hat{F}$, which are non-parametric estimates of the ground truth population $F$.

% For example, let us use the data set $\theta$ and the $\hat{\theta}$ sample data set generated with SRSWOR. The sample data set $\hat{\theta}$ has 20 records in it, and we need a 100 bootstrap sample for the given $\hat{\theta}$. In this case, the number of bootstrap samples (N) is 100. The bootstrap samples are constructed using non-parametric bootstrap \cite{carpenter2000bootstrap}.

\section{Problem Statement} \label{sec:problem}

% - selection query
% - query result based on the tuple result
% - bootstrap sampling based on a sample
% - bootstrap replica calculation for error assessment

\subsection{Selection Query Estimation}

In this research, the considered query formulation, $Q$, is as follows:

\begin{verbatim}
    SELECT Agg(attributes) FROM table WHERE conditions;
\end{verbatim}

The \texttt{Agg} is an analytical aggregation function. In this pilot study, we focus on the \texttt{COUNT} function in the analytic query. A \texttt{COUNT} query can be considered as a simpler case of \texttt{SUM} and \texttt{AVG} queries in AQP. \cite{yu2019cs} introduced a comprehensive AQP framework for common aggregate functions such as \texttt{SUM} and \texttt{AVG}.

\subsection{Query Size Estimation}

To estimate the query $Q$ using approximate query processing, it will first be executed on the sample $S$ of the original table. Each sample tuple $u_i\in S$ will produce a sample query result $y_i$ based on the aggregation function. If the aggregation function is \texttt{COUNT} and the primary key is included in the attribute collection, then $y_i$ will either be 1 if $u_i$ satisfies the selection condition, or 0 otherwise. The query result $Y_s$ on the sample $S$ can be calculated as $Y_s = \sum^n_{i=1} y_i$, where $n=|S|$ or the sample size.

Given the size of the original table $R$ as $N$, the sample fraction $f=n/N$ and the ground truth query result of $Q$ on the original table $R$ can be estimated as 

\begin{equation}\label{eq:est:sample}
    \widehat{Y} = \frac{Y_s}{f}
\end{equation}

\subsection{Error Assessment for Query Estimation}

Equation \ref{eq:est:sample} is an unbiased estimation of the query result of $Q$. The next step is to assess the error produced by the estimation. Bootstrap sampling is a useful tool that can be employed to assess the query estimation based on the given sample query results.

When the aggregation is \texttt{COUNT}, the sample query result $S_Q=\{y_i\}^n_{i=1}$ includes the tuple contribution, $y_i$ either 1 or 0, depending $u_i$ satisfying the query condition or not. We can perform bootstrap sampling on $S_Q$ and generate bootstrap samples $\{\vec{y}_j\}^{B}_{j=1}$ where $B$ is the total number of bootstrap samples. Each $\vec{y}_j=\{y_{j,i}\}^n_{i=1}$ is a bootstrap resample of $S_Q$. 
%each $y_{i,j}$ .

Using Equation \ref{eq:bs:se} on each $\vec{y}_j$, $j=1,...,B$, we obtain the bootstrap replications,

\begin{equation}\label{eq:est:bs-sample}
    \widehat{Y}_j=\frac{Y_{\vec{y}_j}}{f}
\end{equation}

For example, if the aggregation is COUNT, then the estimator is 

\begin{equation}\label{eq:est:bs-sample:count}
    \widehat{Y}_j=\frac{1}{f}\sum^n_{i=1}y_{j,i}
\end{equation}

The bootstrap sampling is repeated for $B$ times, and a collection of bootstrap replications is obtained, denoted by $\widehat{Y}_B=\{\widehat{Y}_j\}^B_{j=1}$. The bootstrap standard deviation is computed as

\begin{equation}\label{eq:std:bs:query}
    \widehat{se}_B=\left[\frac{1}{B-1}\sum^B_{j=1}(\widehat{Y}_j-\overline{\widehat{Y}}_B)^2\right]^{\frac{1}{2}}
\end{equation}
where $\overline{\widehat{Y}}_B$ is the sample mean of all bootstrap replications $\widehat{Y}_B$.

Suppose the significance level is denoted as $\alpha$, where $\alpha$ is a probabilistic value. Common choices of $\alpha$ include 5\% for a 90\% level of significance and 2.5\% for a 95\% level of significance, respectively. The standard method for bootstrap CI (confidence intervals) is calculated as 
\begin{equation}\label{eq:ci:standard}
\left(\widehat{Y}-z^{(1-\alpha)}\cdot \widehat{se}_B, \widehat{Y}+z^{(1-\alpha)}\cdot \widehat{se}_B\right)
\end{equation}
where $\widehat{Y}$ is the query estimation and $z^{(1-\alpha)}$ is the $100(1-\alpha)$th percentile of a standard normal distribution. For example, for 90\% level of significance, $z^{(.95)}$=1.645, and for 95\% level of significance, $z^{(.975)}$=1.960.

\subsection{Quantum-Based Bootstrap Sampling for Error Assessment}

In this research, we would like to investigate how to use quantum computing to accelerate bootstrap sampling for error assessment of query estimations.

The most computationally intensive procedure in bootstrap sampling is to generate a large number of bootstrap samples. Each bootstrap sample consists of random samples with replacements from a given sample of data. For the problem of approximate query processing, the sample data shall be the tuple sample results in $S_Q=\{y_i\}^n_{i=1}$. 

The technical questions include the following:
\begin{enumerate}
    \item \textit{Quantum bootstrap resampling}: How to employ the quantum method to generate random samples with replacement given a set of tuple sample results. This will involve generating random sampling IDs with replacement using quantum computing.
    \item \textit{Quantum bootstrap replication}: How to compute bootstrap replications based on the quantum bootstrap samples generated. We need to translate the randomly sampled tuple IDs into the query tuple results.

    %After using quantum bootstrap sampling, how to use the obtained results to calculate the bootstrap standard deviation and compute the confidence interval to assess the approximate query processing.
\end{enumerate}

\section{Quantum Bootstrap Sampling} \label{sec:methodology}

% - quantum resampling with replacement
% - algorithm description
% - quantum circuit design

This research introduces a \textit{hybrid} framework, including both classical and quantum computing, to accelerate bootstrap sampling for AQP error assessment. Two of the internal processes will employ quantum computing techniques, including the bootstrap random index generation for bootstrap resampling and the query tuple result computation for bootstrap replication computation. After obtaining many bootstrap replications, the total bootstrap standard deviation can be efficiently computed on a classical computer by Equation \ref{eq:bs:se}.

\subsection{Quantum Resampler}

We employ quantum superpositions to produce resampled tuple IDs with replacement from a given set of sample tuple results $S_Q$. Given an $n$-qubit quantum system and a pure state of $\ket{0}^{\otimes n}$, a superposition can be produced by using the Hadamard gates $H^{\otimes n}$, which will be $\psi_1=(H\ket{0})^{\otimes n}$. Each qubit of $\psi_1$ is $H\ket{0} = \frac{1}{\sqrt{2}}(\ket{0}+\ket{1})$, or a one-qubit superposition, that can be measured either $\ket{0}$ or $\ket{1}$ each with a 50\% probability, respectively. 

The $n$-qubit superposition bit string converted to decimal will represent random numbers for resampled tuple IDs with the same probability. In addition, the quantum superpositions will not inhibit repetitions of random index numbers. Observed in different quantum measurements, the same index number generated from the superpositions can be repeated. This enables \textit{sampling with replacement} of the tuples in the given sample dataset. Therefore, the quantum superposition can generate uniform random samples with replacement from the given sample data, which can be used for bootstrap sampling.

With the bootstrap sample index number $\ket{i}$ generated, we employ QRAM \cite{arunachalam2015robustness} to convert $\ket{i}$ to the tuple sample results, $y_i$, in $S_Q$. QRAM can store $S_Q$ into qubit format and encode an index number $\ket{i}$ to the tuple data $\ket{y_i}$ in $S_Q$. A common QRAM implementation is using the Bucket-Bridge approach\cite{giovannetti2008quantum}. Our focus in this work is to use QRAM for tuple retrieval.

After passing the $\ket{i}$ through the QRAM we get $\psi_2 = 2^{-\frac{n}{2}}\sum^n_{i=1} \ket{i}\ket{y_i}$ which represents the tuple qubits in a quantum bootstrap sample $B_i$. Assuming we measure after QRAM, the measurement of $\psi_2$ will be an equal chance of $\ket{y_i}$ in $S_Q$, which is encoded in quantum. The module of quantum resampling can be considered as one circuit named \textit{Quantum reSAmpler} (\textit{QSA}).

\subsection{Quantum Bootstrap Replication}

In order to obtain a bootstrap resample $\vec{y}_j=\{y_{j,i}\}^n_{i=1}$, we need to perform $n$ times of resampling the same size as the sample $S_Q$. We can either repeatedly run the QSA for $n$ times or create $n$ QSA in parallel. For a simpler design of quantum circuits, we use the parallel QSA design, which comprises $n$ QSA that can simultaneously create $n$ of $\ket{y_{j,i}}$'s.

To sum the tuple result values in a quantum bootstrap sample, we employ a quantum counter \cite{heidari2017novel,jiang2024using}, denoted by $QC$, which can perform counting on qubit inputs. In this research, the aggregation function is \texttt{COUNT} and the tuple result $y_{j,i}$ is either 1 if the tuple satisfies the filtering condition or 0 otherwise. The quantum counter $QC$ will count the sum of the corresponding value $y_{j,i}$ in a quantum bootstrap sample encoded in $\ket{y_{j,i}}$, which can be either $\ket{0}$ or $\ket{1}$. The quantum counter comprises control registers and counter registers. Each time there is a $\ket{1}$ input to the control registers, the counter registers will increment by $\ket{1}$, which represents a binary counting number in Least Significant Binary (LSB) form.

After passing $\psi_2$ through $QC$, we will obtain the assembled state of $\psi_3 = \sum^n_{i=1} \alpha_i \ket{i} \ket{y_{j,i}} |\#\{y_{j,i}\in \vec{y}_j | y_{j,i}=1\}|$, where $\alpha_i$ is the amplitude of the quantum bootstrap sample $\vec{y}_j$ generated by the quantum resampler, and $\sum^n_{i=1} \alpha^2_i=1$. After measuring $\psi_3$ at the end of the quantum circuit and translating the quantum value to binary using LSB, we obtain the classical sum of $y_{j,i}$ in $\vec{y}_j$, which is $Y_{\vec{y}_j}$. Using a classical computer, we can convert $Y_{\vec{y}_j}$ to $\widehat{Y}_j$ by Eq \ref{eq:est:bs-sample}, $\widehat{Y}_j=Y_{\vec{y}_j}/f$, where $f=n/N$ is the sampling ratio.

Algorithm \ref{alg:qbs} demonstrates the quantum bootstrap sampling procedure. We can repeat the procedure many times as required and accumulate a collection of classical values of bootstrap replications $Y_{\vec{y}_j}$ or $Y_{B_i}$. We can employ a classical computer to calculate the bootstrap standard deviation and the confidence interval.

Algorithm \ref{alg:quantum_counter} demonstrates the construction of a quantum counter. The arguments $p$ and $q$ determine the number of control qubits and counter qubits. The control qubits can only be either $\ket{0}$ or $\ket{1}$ state. The counter qubits are initialized as $\ket{0}$ state. The designed quantum counter accumulates how many control qubits are in the state $\ket{1}$. Given $p$ control qubits, $q=\log_2(p+1)$ counter qubits will be sufficient to count all possible inputs. 

\begin{algorithm}[t]
\caption{Quantum Bootstrap Sampling}
\label{alg:qbs}
\textbf{Input:} $S_Q=\{y_i\}^n_{i=1}$: Sample tuple results\\
\textbf{Output:} $\widehat{Y}_{B_i}$: A bootstrap replication of query estimation
\begin{algorithmic}[1]
\State $\psi_1 \gets (H\ket{0})^{\otimes n}$ \Comment{Generate a superposition using Hadamard gates}
\State $\psi_2 \gets$ \text{\textsc{QRAM}}($\psi_1$) \Comment{Pass $\psi_1$ through the QRAM}
\State $\psi_3 \gets$ \text{\textsc{Quantum\_Counter}}($\psi_2$, $n$, $\log_2(n+1)$) \Comment{Pass $\psi_2$ through the quantum counter}
\State $Y_{B_i} \gets $ Measuring $\psi_3$ \Comment{Measure a quantum output}
\State \textbf{return} $\widehat{Y}_{B_i}\gets\frac{Y_{B_i}}{f}$ \Comment{Compute on a classical computer} 
\end{algorithmic}
\end{algorithm}

\begin{algorithm}[t]
\caption{\textsc{Quantum\_Counter}($\cdot, p,q$)}
\label{alg:quantum_counter}
\textbf{Input:} $p$: number control qubits; $q$: a factor number of count qubits\\
\textbf{Output:} QC: a quantum circuit of the quantum counter
\begin{algorithmic}[1]
\State QC $\gets$ a quantum circuit with $p$ control qubits $s_0, \dots, s_{p-1}$ and $q$ counter qubits $u_{0}, \dots, u_{q-1}$ 
\For{$i \gets p-1$ \textbf{to} 0}
    \For{$j \gets q-1$ \textbf{to} 0}
    %\For{$u_j \in s_i$ with ascending $j$}
        \State Add an MCX gate with $s_i$, $u_{0}$,..., $u_{j-1}$ as control and $u_{j}$ as the target qubit
        %\State Add a CX gate \Comment{take $s_i$ as control and $u_{j}$ as target}
    \EndFor
\EndFor
\State \textbf{return} QC \Comment{generate quantum circuit}
\end{algorithmic}
\end{algorithm}

% \begin{algorithm}[H]
% \caption{\textsc{Quantum\_Counter}$(p, q)$}
% \label{alg:quantum_counter}
% \textbf{Input:} $p$: number control qubits; $q$: a factor number of count qubits\\
% \textbf{Output:} QC: a quantum circuit of the quantum counter
% \begin{algorithmic}[1]
% \State $c \gets \lceil \log_2(p + 1) \rceil$
% \State QC $\gets$ a quantum circuit with $p$ control qubits $s_0, \dots, s_{p-1}$ and $q \cdot c$ counter qubits $u_{0,0}, \dots, u_{0,c-1}, \dots, u_{q-1,0}, \dots, u_{q-1,c-1}$ 
% \Statex

% \For{$i \gets 0$ \textbf{to} $p-1$}
%     \For{$u_j \in s_i$ with ascending $j$}
%         \For{$k \gets c - 1$ \textbf{downto} $1$}
%             \State Add an MCX gate \Comment{take $s_i$ and $u_{j,0}, \dots, u_{j,k-1}$ as controls and $u_{j,k}$ as the target qubit}
%         \EndFor
%         \State Add a CX gate \Comment{take $s_i$ as control and $u_{j,0}$ as target}
%     \EndFor
% \EndFor
% \State \textbf{return} QC \Comment{generate quantum circuit}
% \end{algorithmic}
% \end{algorithm}

\subsection{Quantum Circuit Design}

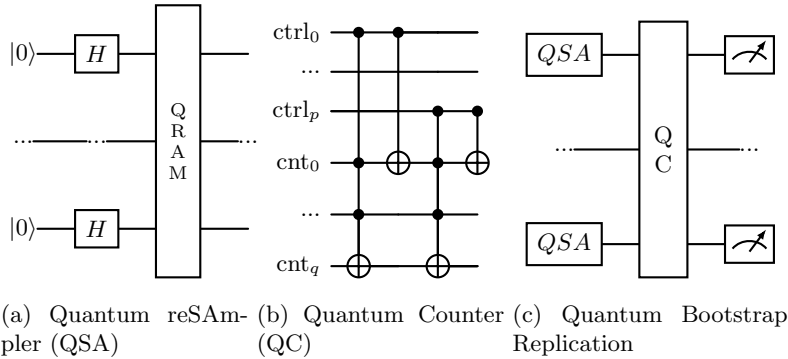
\begin{figure}[t]
    \centering
    \begin{subfigure}[b]{0.27\textwidth}
        \centering        
        \begin{quantikz}
        %\begin{quantikz}[row sep=0.05cm, column sep=0.2cm]
        \ket{0} & \gate{H} & \gate[3]{\scriptsize{\verticaltext{QRAM}}} &  \\
        % \ket{0} & \gate{H} \slice{$\psi_1$} & \gate[3]{\verticaltext{QRAM}} \slice{$\psi_2$} &  \\
        ... & ... &  & ... \\
        \ket{0} & \gate{H} & &
        \end{quantikz}
        \caption{Quantum reSAmpler (QSA)}
        \label{fig:quantum_sampler}
    \end{subfigure}
    \begin{subfigure}[b]{0.27\textwidth}
        \centering
        %\begin{quantikz}[row sep=0.4cm, column sep=0.5cm]
        \begin{quantikz}[row sep=0.45cm, column sep=0.2cm]
        \lstick{$\text{ctrl}_0$} & \ctrl{3} & \ctrl{3} & \qw      & \qw      \\
        \lstick{...} & & & &       \\
        \lstick{$\text{ctrl}_p$} & \qw      & \qw      & \ctrl{3} & \ctrl{1} \\
        \lstick{$\text{cnt}_0$}  & \ctrl{2} & \targ{}  & \ctrl{1} & \targ{}  \\
        \lstick{...} & \ctrl{1} & & \ctrl{1} &  \\
        \lstick{$\text{cnt}_q$}  & \targ{}  & \qw      & \targ{}  & \qw      \qw
        \end{quantikz}
        \caption{Quantum Counter (QC)}
        \label{fig:quantum_counter}
    \end{subfigure}
    \begin{subfigure}[b]{0.3\textwidth}
        \centering        
        \begin{quantikz}[row sep=0.8cm, column sep=0.5cm]
        \gate{QSA} & \gate[3]{\verticaltext{QC}} & \meter{}\\
        % \gate{QSA} & \gate[3]{\verticaltext{QC}} \slice{$\psi_3$} & \meter{}\\
        ... & & ... \\
        \gate{QSA} & &   \meter{}
        \end{quantikz}
        \caption{Quantum Bootstrap Replication}
        \label{fig:quantum_bootstrap}
    \end{subfigure}
\vspace{-6pt}
\caption{Quantum Circuit Design}
\label{fig:quantum_circuit}
\end{figure}

Figure \ref{fig:quantum_circuit} includes the circuit design of the quantum sampling framework. Figure \ref{fig:quantum_sampler} depicts the design of the quantum resampler (QSA). The input states are initialized in $\ket{0}^{\otimes n}$. A superposition is produced after the Hadamard gates to simulate the randomly sampled indices with replacement. After that, the QRAM will translate the random sampling indices to the sample tuple results. 

Figure \ref{fig:quantum_bootstrap} illustrates the overall quantum bootstrap sampling framework. The sampled tuple results from multiple QSA in parallel will be fed into a quantum counter QC. The quantum counter will count the total of non-zero tuple results in each bootstrap resample, which can be measured and converted to a classical bit value in LSB format to produce a bootstrap replication.

% \vspace{-20pt}

% \begin{figure}
%     \centering
%     \begin{quantikz}
%     \lstick{$\text{ctl}_0$} & & \ctrl{5} & \ctrl{3} & \\
%     ... & & & & ... \\
%     \lstick{$\text{ctl}_n$} & & \ctrl{1} & &\\
%     \lstick{$\text{cnt}_1$} &  &    & \targ{} &  \\
%     ... & & & & ... \\
%     \lstick{$\text{cnt}_m$}&  &  \targ{} &  &
%     \end{quantikz}
%     \caption{Simplified Quantum Counter}
%     \label{fig:simplified_quantum_counter}
% \end{figure}

% \vspace{-10pt}

\subsection{Complexity}

Algorithm \ref{alg:qbs} can instantly compute a bootstrap replication after each measurement. The complexity of each measurement is $O(1)$. To generate $M$ bootstrap replications, the total complexity will be $O(M)$.

\subsection{SUM and AVG}

To process \texttt{SUM} and \texttt{AVG} queries, actual numeric values $y_i \in \mathbb{R}$, instead of binary values (0 or 1), can be employed for input sample $S = \{u_1, u_2, ..., u_n\}$. These values will be encoded in QRAM using a fixed-width binary format, for instance, using 5 qubits for values up to 31. The QRAM can map an address state $\ket{i}$ to a data value $\ket{y_i}$ using Toffoli and X gates. For example, address $\ket{001}$ can be mapped to data $\ket{10100}$ for $y_1 = 20$. 
% This creates the state: $|\psi_2\rangle = \sum_{i=0}^{n-1} \alpha_i |i\rangle |y_i\rangle$.
To accumulate QRAM outputs, we can employ the quantum ripple-carry adder circuit \cite{cuccaro2004new} that can add QRAM outputs to an accumulated sum.

% quantikz tutorial
% https://mirrors.ibiblio.org/pub/mirrors/CTAN/graphics/pgf/contrib/quantikz/quantikz.pdf

% \begin{figure}
%     \centering
%     \begin{quantikz}
%     \ket{0} & \gate{H} & \gate[3]{QR} & \gate[3]{QC}  & \meter{}\\
%     %\ket{0} & \gate{H} & & &  \\
%     %\wave &&&& \\
%     ... &...&&& ... \\
%     \ket{0} & \gate{H} & & & \meter{}
%     %& & & \underbrace{}_{\text{\tiny optional}} & 
%     \end{quantikz}
%     \caption{Quantum Circuit Design}
%     \label{fig:quantum-circuit}
% \end{figure}

% \begin{figure}
%     \centering
%     \begin{quantikz}
%     \ket{0} & \gate{H} & \gate[3]{QR}   \\
%     %\ket{0} & \gate{H} & & &  \\
%     %\wave &&&& \\
%     ... &...&  \\
%     \ket{0} & \gate{H} &   
%     %& & & \underbrace{}_{\text{\tiny optional}} & 
%     \end{quantikz}
%     \caption{Quantum Sampler (SA)}
%     \label{fig:quantum-sampler}
% \end{figure}

% \begin{figure}
%     \centering
%     \begin{quantikz}
%     \gate{SA} & \gate[3]{QC} & \meter{}\\
%     %\ket{0} & \gate{H} & & &  \\
%     %\wave &&&& \\
%     ... &...& ... \\
%     \gate{SA} & &   \meter{}
%     %& & & \underbrace{}_{\text{\tiny optional}} & 
%     \end{quantikz}
%     \caption{Quantum Circuit Design}
%     \label{fig:quantum-circuit}
% \end{figure}

\section{Experiment} 
\label{sec:experiment}

\subsection{Experiment Design}

The quantum bootstrap sampling includes two procedures, namely quantum bootstrap resampling and quantum counting. We designed two experiments to simulate the two procedures. We used the IBM Qiskit \cite{javadi2024quantum}, a Python-based high-level quantum design software, to implement the proposed quantum circuit and conduct simulated experiments. Because of the current limitation in quantum computing, the experiments involved a maximum of three qubits, which can process bootstrap sampling of $n=2^3=8$ different tuples. The source code of the experiments is available on GitHub\footnote{Experiment code: \url{https://github.com/YSU-Data-Lab/quantum_bootstrap.git}}.

% In practice, AQP systems usually employs a tiny sampling ratio, such as less than 1\%; therefore, the total tple result size $n$ are also be small.

%%------------------------------------------

\subsection{Quantum Resampling Test}

\begin{figure}[t]
    \centering
    \includegraphics[keepaspectratio, width=1\textwidth]{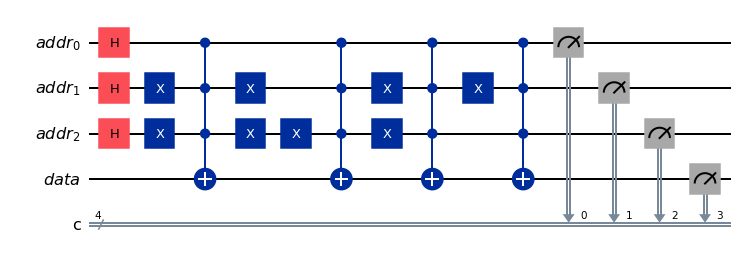}
    \caption{An Implemented QRAM Circuit for Bootstrap Resampling}
    %\caption{Quantum circuit for QRAM encoding showing Hadamard superposition, conditional X encoding gates, multi-controlled Toffoli gates, and measurements}
    \label{fig:qram_implemented}
\end{figure}

Figure \ref{fig:qram_implemented} shows a circuit implementation for quantum sampling with replacement. We initialize our experiment with three qubits in the $\ket{0}$ state. In this circuit, each $\ket{0}$ goes through the Hadamard gate, resulting in the superposition of $\frac{1}{\sqrt{2}}(\ket{0} + \ket{1})$ that generates all possible paths simultaneously. The superposition undergoes a QRAM, which encodes the superposition into a quantum memory address and retrieves the data state from that address.

% \begin{align*}
% \sum_{i=0}^{N-1} \alpha_i \ket{i}_r 
% &\xrightarrow{\text{QRAM}} 
% \sum_{i=0}^{N-1} \alpha_i \ket{i}_r \ket{X_i}
% \end{align*}

For demonstration, the QRAM in this experiment is initialized using a static array where each odd address has a value of 1, and each even address has a value of 0, namely Array = $[0, 1, 0, 1, 0, 1, 0, 1, 0]$. In the AQP application for a \texttt{COUNT} query, value 0 means the tuple result is 0 or not selected, and 1 means the tuple result is 1 or selected. The values of 0 and 1 will be encoded as $\ket{0}$ and $\ket{1}$, respectively. 
% In the QRAM circuit implemented, when we have an address register as $\ket{0}$, nothing changes, no qubits will be flipped, which tells us the indices are even and vice versa. 

% The superposition states of the address along with the data values coming out of the QRAM traverse through the Quantum counter \ref{fig:quantum_bootstrap}. We count the number of data values that are possibly 1 and calculate the total sum using our designed quantum counter \ref{fig:quantum_counter} through measurement.

% We have implemented a simple QRAM encoding circuit \ref{fig:sampler_qram} using IBM Qiskit with the objective of resampling with replacement of all possible address register states. We resampled quantum state to approximate classical statistical patterns, illustrating Quantum Bootstrap sampling. 

The QRAM implemented in this experiment leverages insights from the Bucket-Brigade scheme\cite{giovannetti2008quantum}, where we employ three qubits to demonstrate how classical data associated with a register's address is encoded and retrieved through the properties of superposition and entanglement.

The procedure for finding the data qubit values (sample tuple results $y_i$) corresponding to the superposition address register states is as follows:

\begin{enumerate}
    % \item We have built our circuit using libraries QuantumRegister and ClassicalRegister to define quantum and classical memory. We used AerSimulator to simulate state vectors and measure the results of resampling with replacement. 
    % \item We define a classical data dictionary that represents the QRAM. Here, each address is mapped to a binary value representation from 0 to 7, as we have initialized with 3 qubits. 
    % We did the mapping by: \texttt{qram\_data} = $\{ i | i \mod 2 \}$
    % Here, the qram\_data results address registers with odd indices containing data values 1 and even indices having data value 0.\\
    \item To set up the register, we employ three address qubits (addr$_i$) representing $2^3$=8 memory locations. We have one data qubit (data) that stores the output value corresponding to the selected address. A qubits are initialized as $\ket{0}$ states. A classical register (c) of size four is used to store the measurement results from both the address and data qubits.
    \item Using three Hadamard gates, the address qubits are converted into the superposition states, generating all possible memory addresses in parallel.
    \item We employ Toffoli gates (also known as Multi-Controlled NOT gate or MCX gate) \cite{nielsen2010quantum} to flip the data qubit according to the states of address qubits. X (NOT) gates are employed to control the address qubits to match the bit-string pattern in the data array for the QRAM. For example, when the address qubits after Hadamard gates are $\ket{000}$, the data qubit shall be measured as $\ket{0}$; whereas for address qubits $\ket{001}$, the data qubit shall be $\ket{1}$.
\end{enumerate}

In this experiment, we have executed 1024 shots for generating samples with replacement using Qiskit's AerSimulator \cite{javadi2024quantum}. Measurements on all addresses and data qubits have been performed accordingly. Due to Qiskit's little-endian output generation format, we reverse the bit ordering of address qubits to make it human-readable.

Table \ref{tab:qram_results} shows all possible QRAM inputs and outputs. Each row correlates to a possible state from sampling measurement results. When the address is odd, the data qubit returns 1 which is consistent with the encoding mechanism.

\begin{figure}[t]
    \centering
    \includegraphics[width=.7\textwidth, keepaspectratio]{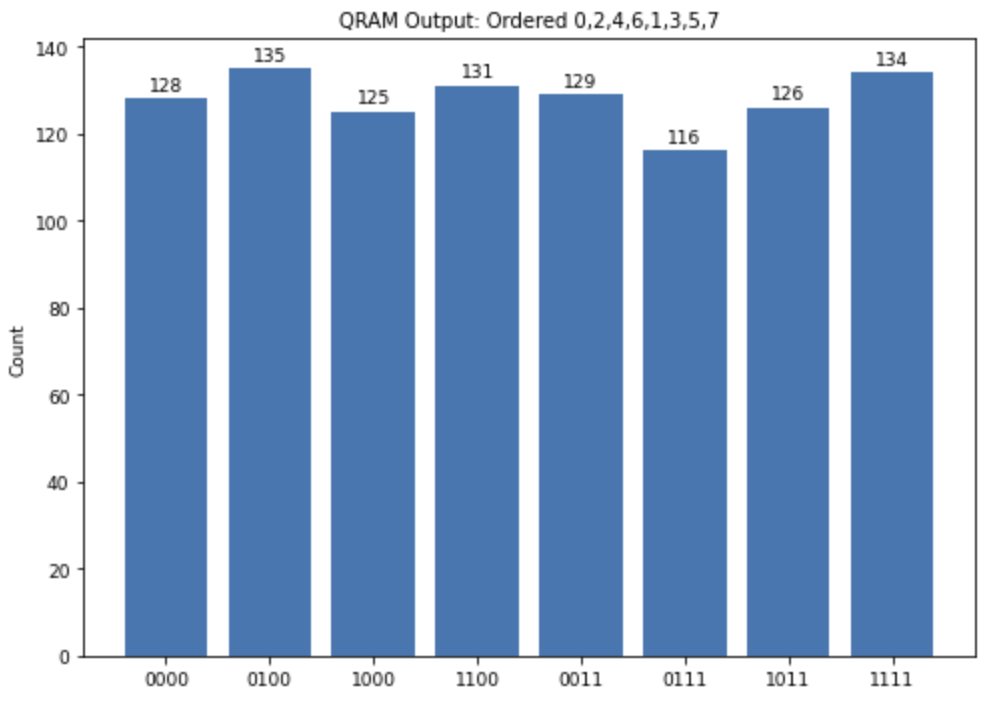}
    \caption{A Distribution of QRAM Output States}
    \label{fig:qram_histogram}
\end{figure}

Figure \ref{fig:qram_histogram} depicts the frequency of each quantum state being sampled. After 1024 experiment shots, the data qubit value having 1 (odd addresses) has similar frequencies to even addresses. It verifies that the functionalities of superposition are intact.

\begin{table}[t]
\centering
\caption{Measured QRAM Output States}
\label{tab:qram_results}
\begin{tabular}{|c|c|c|c|}
\hline
\textbf{Address (Binary)} & \textbf{Address (Decimal)} & \textbf{Data Qubit} & \textbf{Count} \\
\hline
000 & 0 & 0 & 128 \\
010 & 2 & 0 & 135 \\
100 & 4 & 0 & 125 \\
110 & 6 & 0 & 131 \\
\hline
001 & 1 & 1 & 129 \\
011 & 3 & 1 & 116 \\
101 & 5 & 1 & 126 \\
111 & 7 & 1 & 134 \\
\hline
\end{tabular}
\end{table}

%%------------------------------------------

\subsection{Quantum Counter Test}

Figure \ref{fig:quantum_counter} depicts a simplified quantum counter circuit to count the number of qubits in $\ket{1}$ state from the output of a quantum resampler. Our implementation is inspired by Jiang et. al. \cite{jiang2024using}. The output count result is stored in binary format, representing a ripple-carry adder-like binary counter. The output number represents the total of tuple results of 1, or selected, in the \texttt{COUNT} query, which can be used to compute the bootstrap replication. 

The setup of the quantum counter circuit is as follows. In the previous test, we employed three qubits representing the sample data size of $n=8$. 
% We simulate when eight quantum resamplers simultaneously feed the quantum counter with sample tuple results encoded either in $\ket{0}$ or $\ket{1}$. 
In the implemented quantum counter, the number of control qubits is eight and the number of counter qubits is $4=\log_{2}(8+1)$. The function of the quantum counter is to count the total of $\ket{1}$ states in the control qubits. The control qubits are also initialized as $\ket{0}$ and undergo the Hadamard gates to generate a superposition to simulate all possible states of the control qubits. 

Initially, all counter qubits are initialized as $\ket{0}$. The counter qubits are connected to each control qubit using Toffoli and control-NOT gates. The numbers in the counter qubits are represented in the LSB format. For each $\ket{1}$ state in a control bit, the value in the counter bits will be incremented by 1. To simulate the quantum counter for each quantum bootstrap sample, we run the quantum counter circuit using the AerSimulator with one shot each time.

Table \ref{tab:quantum_counter_sample} depicts an example quantum counter output during one simulation. The input control qubits are $\ket{00011111}$, namely the control qubits 0 to 4 are in the $\ket{1}$ state. The output of the quantum counter, or the counter qubits, is $0101_2$ = 5. Therefore, the quantum counter correctly encoded the total $\ket{1}$'s in the control qubits into the counter qubits.

\begin{table}[t]
\centering
\caption{One Result of Quantum Counter Simulation}
\begin{tabular}{|l|l|}
\hline
\textbf{Description} & \textbf{Value} \\
\hline
Full bitstring measured & $\ket{010100011111}$ (last bit to first bit) \\
Control bits measured   & $\ket{00011111}$ (5 ones) \\
Counter bits  measured & $\ket{0101}$ (binary of value 5) \\
\hline
\end{tabular}
\label{tab:quantum_counter_sample}
\end{table}

\begin{figure}[t]
    \centering
    \includegraphics[width=1\textwidth, keepaspectratio]{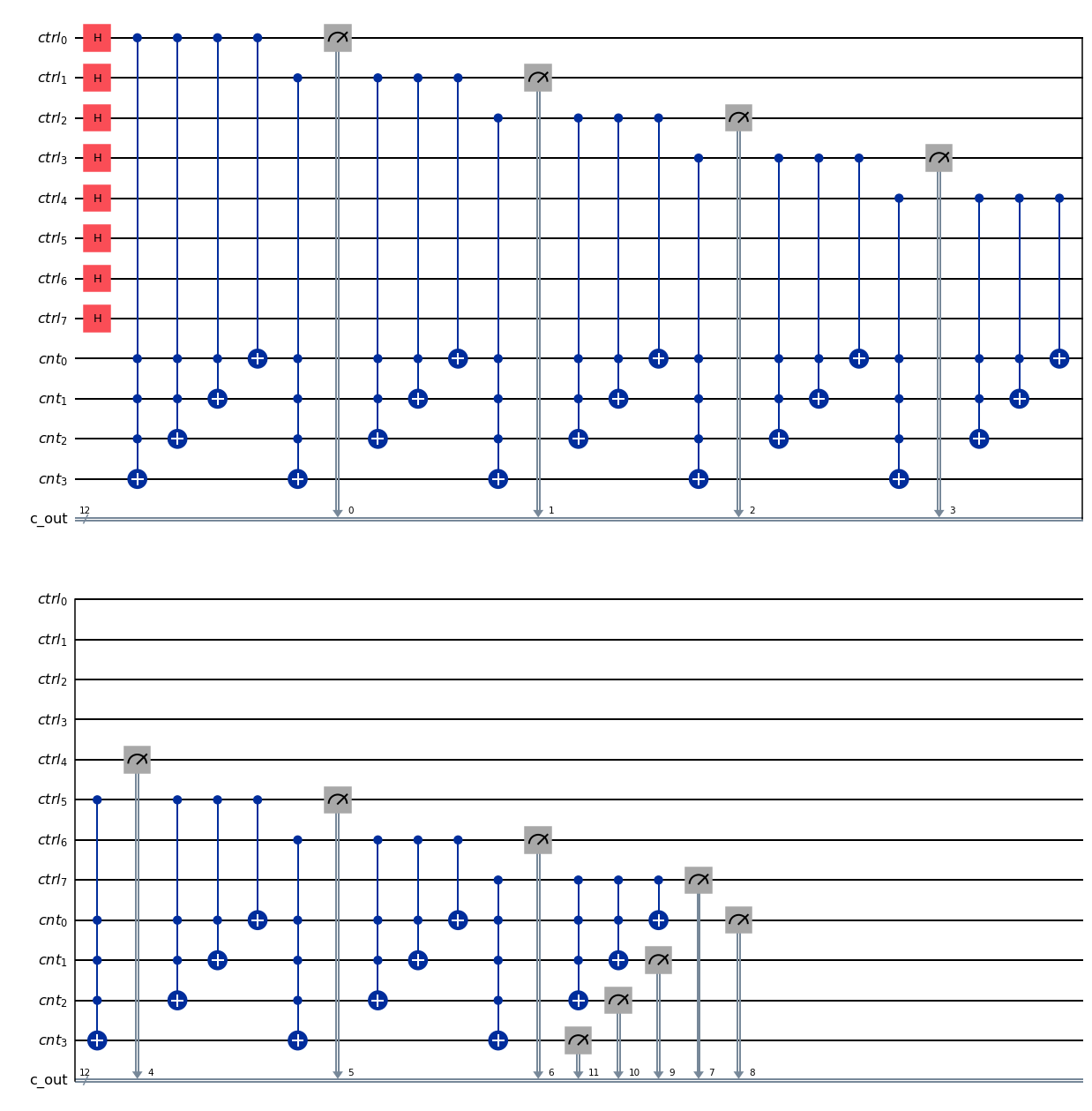}
    %\caption{A Simplified Quantum Counter Circuit for Sample Size $n\leq 8$}
    \caption{A Simplified Quantum Counter to Count Qubits ($n\leq 8$) in $\ket{1}$ State}
    \label{fig:quantum_counter}
\end{figure}

\section{Related Work} \label{sec:related-work}

% Yu add bootstrap sampling works

Wu et. al. \cite{wu2024quantum} proposed a hybrid classical-quantum sampling framework known as an amplifier for sampling-based approximate query processing (AQP) to identify normal and rare groups. It amplifies the signals of the highly selective and skewed distributed rare groups iteratively by implementing Grover's diffusion operator and QRAM. Employing a diffusion operator, the scheme demonstrated the power of quadratic acceleration $O(\sqrt{N})$ to achieve balanced sampling across various data categories. The classical sampling approach categorizes a set of normal and rare groups, whereas the quantum sampling method was used to amplify the rare groups in \cite{wu2024quantum}. The classical sampling-based AQP is considered unsuitable for Group-By queries with high selectivity. Quantum computing can enhance a specific desired quantum state through amplitude amplification without the necessity of data pre-processing. This research has conducted several experiments where it is observed that, with the increased number of amplification iterations, the probability of obtaining desired tuples also increased remarkably. Oracle was used to mark the index 0 of the desired tuple, and the diffusion operator was used to amplify its probability and spread the quantum state around the mean. Finally, using QRAM, they encoded the quantum state $\ket{i}$ to $\ket{t_i}$ according to the address mapping. This work was mainly on addressing the single-table sampling problem for integer values instead of strings.

\cite{wocjan2008speedup,wild2021quantum,singh2023proof,lund2017quantum} studied optimization of sampling for AQP. However, how a quantum algorithm helps optimize bootstrap sampling for faster AQP of large-scale databases is still understudied.

Phalak et. al. \cite{phalak2023quantumrandomaccessmemory} compared various types of QRAM (Bucket-Brigade, Fan-Out, Flip-Flop, EQGAN, Approximate PQC) with classical RAM, focusing on its explanatory perspective, advantages, and significance in exponential speedup for algorithms such as Fourier Transform, discrete logarithm, and pattern recognition. When it comes to database searching, discovering differences in elements, faster data retrieval, and storing of classical and quantum data, QRAM can access multiple memory locations at once due to superposition. QRAM uses quantum swap operations and qubits ($\ket{0}$ and $\ket{1}$) for read and write operations, whereas classical RAM uses bits (0 or 1) and read/write signals. For $n$ address registers, bucket-brigade QRAM\cite{giovannetti2008quantum} is efficient due to the $O(n)$ gate activation mechanism in comparison to RAM having $O(2^{n})$. The output of QRAM is the data state of a specific address state and can be retrieved by performing entanglement operations using SWAP or CNOT gates. The quantum switches in this type of QRAM have three states $\ket{.}$ or waiting state, $\ket{0}$, and $\ket{1}$. 
%Classical RAM is inefficient in storing quantum states as it destroys the superposition of states and collapses the wave function when measurement takes place. 
For Bucket-Brigade QRAM, we can sequentially access the data states from the most significant bit to the least significant bit. Additionally, Fanout QRAM\cite{park2019circuit} uses qubits that control exponential $O(2^n)$ switches for single and superposition of all the addresses. Furthermore, Flip-Flop QRAM uses a multi-controlled rotation gate to store data in the address register qubit. EQGAN\cite{niu2022entangling} is an entanglement-based model that optimizes training the data on a Quantum Neural Network (QNN)\cite{abbas2021power} having a classification accuracy of around 65\%. Approximate PQC-based QRAM\cite{phalak2022approximate} can store complex image data and binary data in sequential order. Loading images from QRAM and sending them to QNN is way faster than without QRAM. Qudit-based memory architectures eliminate the need for ancillary qubits, which helps in reducing hardware overhead. 

% Despite having some parallel processing computational advantages, there are major challenges in terms of scalability of increased memory allocation, noise resilience, error correction, and redundancy schemes of QRAM. In our proposed algorithm, we have used basic Bucket brigade structured QRAM to access and store the data attached to the specific address register of superposed states. 

Jiang et. al. in \cite{jiang2024using} introduced a simplified quantum counter to solve NP-hard exact cover problem, which reduces the number of quantum gates by $(4mb - 4m) [\pi/4 \sqrt{N/M}]$ and circuit depth by $2mb[\pi/4 \sqrt{N/M}]$, where $b$ is counting qubits, $m$ is quantum counters, $N$ is the total number of possible input instances and $M$ is the number of target inputs (or solutions). One question is to find the number of solutions $M$. To address this challenge, this study utilizes the quantum counting algorithm to estimate the number of solutions to the given problem. The quantum counting algorithm depends on quantum phase estimation to derive $\theta$ of the Grover iterator to find $M$. 
Quantum Fourier Transform (QFT) was utilized to transform states to the computational basis of $\ket{0}$ and $\ket{1}$ for measuring counting qubits. $\theta$ can be found by these measurement results. According to this work, having $\theta$ and $N$, the number of solutions $M$ can be found by $M = N\sin^2(\theta/2)$. In the proposed quantum circuit, one quantum counter and one inverse quantum counter for resetting were employed. There are two types of qubit registers, namely controller and counter, in this design. The counter acts as an accumulator, and its value increases by 1 when the control qubit is $\ket{1}$ to determine how many times each element in the universal set is covered by sets in the collection. The oracle of the algorithm uses the MCX gate to control counter qubits and X gates were used to flip the ancilla qubit if every counter gives a value of 1. 
% They have conducted 4 experiments using IBM Qiskit packages and Aer Simulator, for correctness validation of ECP, achieving different output results. 

% If a quantum state is expressed as $\ket{\psi} = \sum_i ai \ket{i}$, measuring this quantum state entails a probability of $a_i^2$ for observing the state $\ket{i}$. They have defined their examples for SUM, AVG and COUNT.
% QRAM was implemented using the Bucket-Brigade technique. Bucket-Brigade QRAM can reduce the access cost to O(log N). The authors have worked with IntegerComparator quantum gate to compare and find out which tuple satisfies the condition and if it passed, then it flips the auxiliary qubit. Multi-controlled X gate was utilized along with auxiliary qubits for phase inversion. For the queries, each query was run 100 times which helped to find out average performance metrics. Their algorithm achieves accuracy quite similar to or even better than stratified sampling with pre-clustering for rare group. Quantum sampling cost is roughly $O(\sqrt{N × NP})$ which is notably low.

% However, their work is not closely related to our research paper as we are working specifically with bootstrap sampling.

% The two prominent quantum circuits we are going to use are QRAM and the quantum counter. In various previous studies, the functionalities of QRAM and quantum counter have been depicted nicely.

\section{Conclusion and Future Work} \label{sec:conclusion}

In this work, we introduced quantum bootstrap sampling, which can be used to accelerate error assessment for AQP query estimations. The framework begins by producing indices for random sampling with replacement. After that, we employed a QRAM and a quantum counter to produce the bootstrap replication. We implemented the prototypes of quantum circuits, including both the quantum resampler and the quantum counter, using IBM Qiskit. The experiments showed that bootstrap resamples can be instantly generated based on testing data. The implemented quantum counter was observed functioning correctly given inputs from the quantum resamplers. In the future, we will further investigate using quantum bootstrap sampling for more types of aggregate functions, such as SUM and AVG. We plan to design experiments using more complicated real datasets.

% In the future, we will implement the quantum circuit on a quantum computer, such as IBMQ, and conduct experiments on test datasets. 

\section{Acknowledgement}
\label{sec:acknowledgement}

This research is partially supported by the Research Advancement Grant, Research Professorship Award, and Student Small Grant at Youngstown State University.

%---bibliography---
\bibliographystyle{template/splncs04}
\bibliography{references/aqp,references/bs,references/my-pub,references/qc,references/citations}

\begin{thebibliography}{10}
\providecommand{\url}[1]{\texttt{#1}}
\providecommand{\urlprefix}{URL }
\providecommand{\doi}[1]{https://doi.org/#1}

\bibitem{abbas2021power}
Abbas, A., Sutter, D., Zoufal, C., Lucchi, A., Figalli, A., Woerner, S.: The
  power of quantum neural networks. Nature Computational Science
  \textbf{1}(6),  403--409 (2021)

\bibitem{arunachalam2015robustness}
Arunachalam, S., Gheorghiu, V., Jochym-O'Connor, T., Mosca, M., Srinivasan,
  P.V.: On the robustness of bucket brigade quantum ram. New Journal of Physics
   \textbf{17}(12),  123010 (2015)

\bibitem{cuccaro2004new}
Cuccaro, S.A., Draper, T.G., Kutin, S.A., Moulton, D.P.: A new quantum
  ripple-carry addition circuit (2004),
  \url{https://arxiv.org/abs/quant-ph/0410184}

\bibitem{efron94bootstrap}
Efron, B., Tibshirani, R.J.: An introduction to the bootstrap. CRC press (1994)

\bibitem{giovannetti2008quantum}
Giovannetti, V., Lloyd, S., Maccone, L.: Quantum random access memory. Physical
  review letters  \textbf{100}(16),  160501 (2008)

\bibitem{heidari2017novel}
Heidari, S., Farzadnia, E.: A novel quantum lsb-based steganography method
  using the gray code for colored quantum images. Quantum Information
  Processing  \textbf{16}(10), ~242 (2017)

\bibitem{javadi2024quantum}
Javadi-Abhari, A., Treinish, M., Krsulich, K., Wood, C.J., Lishman, J., Gacon,
  J., Martiel, S., Nation, P.D., Bishop, L.S., Cross, A.W., et~al.: Quantum
  computing with qiskit. arXiv preprint arXiv:2405.08810  (2024)

\bibitem{jiang2024using}
Jiang, J.R., Wang, Y.J.: Using a simplified quantum counter to implement
  quantum circuits based on grover's algorithm to tackle the exact cover
  problem. Mathematics  \textbf{13}(1), ~90 (2024)

\bibitem{li-tods2019}
Li, F., Wu, B., Yi, K., Zhao, Z.: Wander join and xdb: Online aggregation via
  random walks. ACM Trans. Database Syst.  \textbf{44},  2:1--2:41 (1 2019)

\bibitem{Ling1999}
Ling, Y., Sun, W., Rishe, N.D., Xiang, X.: A hybrid estimator for selectivity
  estimation. IEEE Transactions on Knowledge and Data Engineering  \textbf{11},
   338--354 (1999)

\bibitem{encyc-aqp}
Liu, Q.: Approximate query processing. In: LIU, L., {\"O}ZSU, M.T. (eds.)
  Encyclopedia of Database Systems, pp. 113--119. Springer US (2009)

\bibitem{lund2017quantum}
Lund, A.P., Bremner, M.J., Ralph, T.C.: Quantum sampling problems,
  bosonsampling and quantum supremacy. npj Quantum Information  \textbf{3}(1),
  ~15 (2017)

\bibitem{nielsen2010quantum}
Nielsen, M.A., Chuang, I.L.: Quantum computation and quantum information.
  Cambridge university press (2010)

\bibitem{niu2022entangling}
Niu, M.Y., Zlokapa, A., Broughton, M., Boixo, S., Mohseni, M., Smelyanskyi, V.,
  Neven, H.: Entangling quantum generative adversarial networks. Physical
  Review Letters  \textbf{128}(22),  220505 (2022)

\bibitem{park2019circuit}
Park, D.K., Petruccione, F., Rhee, J.K.K.: Circuit-based quantum random access
  memory for classical data. Scientific reports  \textbf{9}(1), ~3949 (2019)

\bibitem{phalak2023quantumrandomaccessmemory}
Phalak, K., Chatterjee, A., Ghosh, S.: Quantum random access memory for dummies
  (2023), \url{https://arxiv.org/abs/2305.01178}

\bibitem{phalak2022approximate}
Phalak, K., Li, J., Ghosh, S.: Approximate quantum random access memory
  architectures. arXiv preprint arXiv:2210.14804  (2022)

\bibitem{singh2023proof}
Singh, D., Muraleedharan, G., Fu, B., Cheng, C.M., Newton, N.R., Rohde, P.,
  Brennen, G.K.: Proof-of-work consensus by quantum sampling. Quantum Science
  and Technology  (2023)

\bibitem{wang2012quantum}
Wang, Y.: Quantum computation and quantum information  (2012)

\bibitem{wild2021quantum}
Wild, D.S., Sels, D., Pichler, H., Zanoci, C., Lukin, M.D.: Quantum sampling
  algorithms, phase transitions, and computational complexity. Physical Review
  A  \textbf{104}(3),  032602 (2021)

\bibitem{wocjan2008speedup}
Wocjan, P., Abeyesinghe, A.: Speedup via quantum sampling. Physical Review
  A---Atomic, Molecular, and Optical Physics  \textbf{78}(4),  042336 (2008)

\bibitem{wu2024quantum}
Wu, S., Shi, M., Zhang, D., Zhao, J., Yuan, G., Chen, G.: When quantum
  computing meets database: A hybrid sampling framework for approximate query
  processing. IEEE Transactions on Knowledge and Data Engineering  (2024)

\bibitem{yu2022non}
Yu, F., Cal, S., Cheng, E., Kerns, L., Xiong, W.: Non-parametric error
  estimation for $\sigma$-aqp using optimized bootstrap sampling. International
  Journal for Computers \& Their Applications  \textbf{29}(1) (2022)

\bibitem{yu2019cs}
Yu, F., Hou, W.C.: {CS}*: Approximate query processing on big data using
  scalable join correlated sample synopsis. In: 2019 IEEE International
  Conference on Big Data (Big Data). pp. 583--592. IEEE (2019)

\end{thebibliography}
% \bibliographystyle{plain}
% \bibliography{references/citations,references/bootstrap,references/semih_thesis,references/sampling}

\end{document}